\documentclass[aps,twocolumn]{revtex4}

\usepackage{graphicx}

\begin{document}

\title{Global protein function prediction 
in protein-protein interaction networks}

\author{A. Vazquez$^{1,2}$, A. Flammini$^2$, A. Maritan$^{2,3}$ and A. Vespignani$^4$} 

\affiliation{$^1$ Department of Physics, University of Notre Dame, Notre Dame, IN 46556, USA}

\affiliation{$^2$ International School for Advanced Studies (SISSA) and INFM, V. Beirut 2-4, 34014 Trieste, Italy}

\affiliation{$^3$ The Abdus Salam International Centre for Theoretical Physics, P.O. Box 586, 34100 Trieste, Italy}

\affiliation{$^4$ Laboratoire de Physique The©orique (UMR du CNRS 8627), aâtiment 210  Universeté de Paris-Sud 91405 
ORSAY Cedex - France}

\date{\today}

\maketitle

{\bf
The determination of protein functions is one of the most challenging problems of the post-genomic era.  The sequencing
of entire genomes and the possibility to access gene's co-expression patterns has moved the attention from the study of
single proteins or small complexes to that of the entire proteome [1]. In this context, the search for reliable methods
for proteins' function assignment is of uttermost importance.  Previous approaches to deduce the unknown function of a
class of proteins have exploited sequence similarities or clustering of co-regulated genes [2,3], phylogenetic profiles
[4], protein-protein interactions [5,6,7,8], and protein complexes [9,10]. We propose to assign functional classes to
proteins from their network of physical interactions, by minimizing the number of interacting proteins with different
categories. The function assignment is made on a global scale and depends on the entire connectivity pattern of the
protein network.  Multiple functional assignments are made possible as a consequence of the existence of multiple
equivalent solutions. The method is applied to the yeast Saccharomices Cerevisiae protein-protein interaction network
[5]. Robustness is tested in presence of a high percentage of unclassified proteins and under deletion/insertion of
interactions.}

Two hybrid experiments, as that conducted by Uetz et al. [5] , allow to reconstruct the set of physical binary
interactions among a set of proteins of a given proteome. In order to introduce our approach, we visualize the
protein-protein interaction data as a graph whose nodes represent the proteins and edges connect pairs of interacting
proteins [11,12]. The suggestion we want to implement here is that interacting proteins may belong to at least one common
functional class. Therefore the knowledge of the functional classification of a subset of the proteins involved in the
network may lead to an educated guess on the functional classification of the remaining subset of uncharacterized
proteins. In principle, to each protein should assigned one or more functional classes drawn from a set of $F$ possible
classes. $F$ is the total number of functions considered and depends on the functional classification scheme used. Finer
is the definition of function used in the classification scheme and greater is the number $F$. In general, however, the
functional class is known only for a subset of proteins and one faces the problem of assigning a function $\sigma_i$,
chosen among all $F$ possible functions, to each unclassified protein $i$.

\begin{figure}

\centerline{\includegraphics[width=3in]{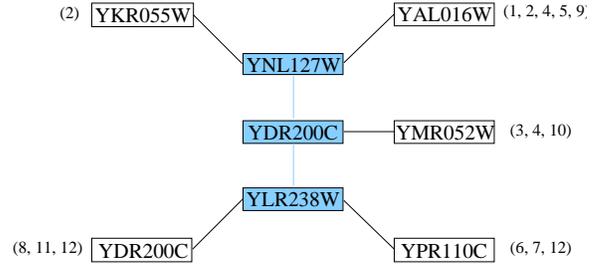}}

\caption{{\bf Illustration of the method.} Sub-graph of the protein-interaction network of the yeast Saccharomices
Cerevisiae. Proteins in gray boxes are unclassified (unknown function) while the others are classified proteins with the
functions within the brackets and labeled according to the following criteria: 1- cell growth; 2- budding, cell polarity
and filament formation; 3- pheromone response, mating-type determination, sex-specific proteins; 4- cell cycle check
point proteins; 5- cytokinesis; 6- rRNA synthesis; 7- tRNA synthesis; 8- transcriptional control; 9- other transcription
activities; 10- other pheromone response activities; 11- stress response; and 12- nuclear organization. Given one of
these proteins of unknown function if we take as a prediction the function that appears more in the neighbor proteins of
known function then we obtain the following classification (from top to bottom) YNL127W (2), YDR200C (3,4,10) and YLR238W
(12). Our method, however, considers also the interactions among unclassified proteins. If we iterate once more the
"majority rule" by taking into account the interactions between the three unclassified proteins, we obtain the
following classification YNL127W (2,4), YDR200C (3,4,10) and YLR238W (12). In this way we find another possible function
for YNL127W. This is actually the spirit of our method, we take advantage of the prediction we are making for proteins of
unknown function and apply a global optimization method.}

\label{fig1}

\end{figure}

Assigning to an unclassified protein the most common function(s) present among the classified interacting proteins, as in
[7,8] ("majority rule" assignment) is rather straightforward. The majority rule relies on the empirical evidence that
70-80\% of interacting proteins pairs share at least one function. In most cases [13], however, few unclassified proteins
have more than one interacting protein with known function. In addition, in these few cases, the interacting proteins
with known functions do not usually have shared functionalities (Fig. 1). In this perspective, the use of the "majority
rule" assignment is rather unsatisfactory since all links between proteins with unknown function are completely
neglected. This implies a very partial use of the knowledge of the protein-protein interaction network. More important,
the final configuration of assigned functions to unclassified proteins ought to be consistent with the rules used to
determine the functions themselves. To an unclassified protein with one or more unclassified partner(s) must be assigned
functions consistently with the functions assigned to the unclassified partners. This points out to a method in which
unknown proteins influence the "majority rule" assignment in a self-consistent way once their functions have been
selected.

Our functional prediction strategy is based on a global optimization principle: a score or energy (see Experimental
protocol section for details) is associated to any given assignment (configuration) of functions for the whole set of
unclassified proteins. The score is lower in configurations that maximize the presence of the same functional annotation
in interacting proteins. The new ingredient is that the contribution to the total score of a given functional assignment
of an unclassified protein is computed as the number of classified and unclassified neighbor proteins with that function.
Hence, the determination of the functions of all unclassified proteins in the network becomes a global optimization
problem and can not be solved on the basis of the local environment only. Finding the optimal function assignment
corresponds to determine the minimal score for the whole network. This corresponds, in statistical mechanics in
minimizing the energy of a Pott's model [14] with non-homogeneous boundary conditions, the latter being represented by
the proteins with known function. The resulting computational problem is frustrated; i.e. it is generally impossible to
satisfy all the constraints imposed by classified proteins on their interacting, unclassified partners. This leads to a
multiplicity of equivalent or nearly equivalent optimal solutions that contain a minimal amount of interacting proteins
with different functions. The existence of multiple solutions allows the objective assignment of multiple functions to
most unclassified proteins (Fig. 1). Depending on the complexity of the underlying graph and on the boundary conditions,
the score minimization represents an hard computational task. In instances of this type, "simulated annealing"
technique (see Experimental protocol section) is an appropriate tool to obtain the optimal solutions. Indeed, the
optimization procedure is repeated several times to account for the non uniqueness of the optimal configurations and a
prediction for the functional classification is finally made by taking those functions that, for each unclassified
protein, occurred more often in the whole set of simulated annealing processes .

We have applied our functional-prediction method to the yeast Saccharomices Cerevisiae protein-protein interaction
network. The interaction data was obtained from Ref. [7] and it contains $N=1826$ proteins with $E=2238$ identified
interactions.  The functional classification was obtained from the MIPS database [15]. The MIPS finer classification
scheme contains $F=424$ functional categories, plus two categories for proteins with no assigned function:  
'CLASSIFICATION NOT YET CLAR-CUT' and 'UNCLASSIFIED PROTEINS'. The data contains $n=441$ proteins in these two last
categories. We used our global optimization method to obtain the functional assignments of all the proteins listed within
these two categories. The complete set of functional assignments can be found as Supplementary Table 1. For each
unclassified protein we report its degree, i.e. number of proteins directly connected to it, and up to three of the most
probable predicted functions as found with our method. We attribute a higher level of "certainty" to those functions
with a higher percentage of occurrence.

\begin{table}
\begin{tabular}{|l|l|l|l|l|}
\hline
$k$ & $n_k$ & MR1 & GO1 & GO2\\
\hline
2 & 328 & 0.40 & 0.46 & 0.61\\
3 & 205 & 0.55 & 0.65 & 0.76\\
4 & 102 & 0.60 & 0.62 & 0.77\\
5 & 72 & 0.58 & 0.66 & 0.86\\
6 & 41 & 0.66 & 0.74 & 0.89\\
7 & 28 & 0.58 & 0.67 & 0.94\\
$k>7$ & 85 & 0.69 & 0.74 & 0.94\\
\hline
\end{tabular}

\caption{{\bf Global optimization vs majority rule}. Comparison of the success rate of the global optimization (GO)
method proposed here and the majority rule (MR). To compute the success rate we assume that a fraction $f_n=0.4$ of the
classified proteins are unclassified and then make functional predictions for them. The success rate is defined as the
probability that the most ranked predicted function is actually a functional classification for the corresponding
protein. Two different levels of functional classification have been used. In the finest level (1) we have taken the
finest classification, containing 424 functional categories. In the coarse-grained level (2) we have taken the highest
level classification (metabolism, energy, cell growth and division, etc.), containing 20 functional categories. We show
the success rate as a function of the number of interacting partners $k$ (as a reference we also show how many proteins
$n_k$ has $k$ interacting partners). The case $k=1$ is not considered since the MR method finds only a trivial
implementation in this case. The comparison of the values for $k\geq2$ clearly indicates that the global optimization
method performs better, with higher percentage of correct predictions. Moreover, the more coarse-grained is the
classification the higher is the success rate. Not surprisingly, adopting a coarser classification scheme leads to an
increasing of the various rate of success (last column) since the number of degrees of freedom the method has do deal
with is drastically reduced. This, of course, has to be balanced with the parallel diminution of the information content
of predictions.}

\end{table}

A fundamental issue concerning protein function predictions is the assessment of the method reliability with respect to
the incomplete knowledge of the interaction network. In order to establish an upper bound to the predictive power of our
method, we ignored the functionality of a finite fraction $f_n$ of the classified protein and then measured the rate of
successful predictions of our method by comparing with their real classification. In this way, one can get a quantitative
estimate of the reliability of our predictions as a function of the amount of available information about the network.
Fig. 2a shows the percentage of successful predictions as a function of the degree of the proteins for different values
of $f_n$ using the finest functional classification scheme available (424 functional classes). In the case of
unclassified proteins with degree larger than 2, a correct prediction can be made between 60\%-70\% of the cases, even
with the loss of a substantial part of the information (up to $f_n=0.4$) and quite independently of the degree of the
protein involved in the prediction. A visual inspection of the test for $f_n=0.4$ can be obtained browsing the
Supplementary Table 2. A quantitative account of the better performance of our method with respect to local optimization
methods is presented in Tab. 1, where we also report predictions obtained with a coarser classification scheme. These
results prompt to a large and stable statistical predictive power of the method also in the presence of reduced
information (larger number of unclassified proteins).

\begin{figure}

\centerline{\includegraphics[width=2.8in]{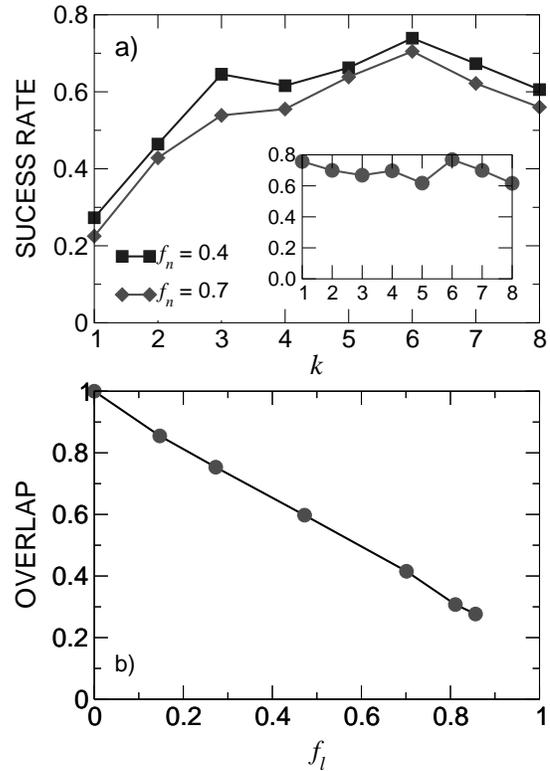}}

\caption{{\bf a)Self-consistent test.} The success rate of our method after a fraction $f_n$ of classified proteins has
been set unclassified. Each point represents the probability that the functional classification of proteins with $k$
interacting partners, defined here as the top ranking for occurrence in the list of putative functions generated by our
method, coincides with their real classification. We report the success rate for the values $f_n=0.4$ and $f_n=0.7$ in
the upper and lower curve, respectively. The prediction quality for poorly connected nodes (degree 1 and 2) decreases to
just 30\% and it is degrading more rapidly than for highly connected proteins. This is a consequence of the fact that the
corresponding proteins occupy a very marginal position where the method cannot take fully advantage of the global
connectivity properties of the graph. In the inset we report the data for $f_n\rightarrow0$, {\it i.e.} when only a
single protein is set unclassified. In this case it is possible to see that even for poorly connected proteins the
methods gives a very good statistical reliability of the corresponding predictions. {\bf b) Tolerance to errors.} The
inset shows the overlap ($\Theta_i(f_l)$ averaged over all unclassified proteins) between the functional predictions
obtained using the original network and another with a degree of dissimilarity $f_l$ defined as the percentage of edges
between proteins couples which are different in the two networks. The analysis of the figure shows that a moderate amount
of misplaced interactions do not preclude a reliable function assignment. Of course larger levels of errors lower the
overlap, signaling that the two networks provide rather different configurations of functional assignment. The curve show
a decreasing linear trend that extrapolated to $f_l=1$ gives a value of the overlap very small (less than 15\% ), as
expected. Still, the extrapolation to $f_l=1$ is somewhat inappropriate since a complete dissimilarity between the
original and the scrambled network is hardly achievable by a random rewiring, and therefore unjustified in the present
context.}

\end{figure}

A second test concerns the presence of errors in the topology of the protein network. It is known that protein-protein
interactions data obtained from two hybrid experiments contain an amount of false positives and negatives and these could
in principle alter sensibly the quality of predictions by providing a spurious connectivity to the network (false or
missing edges). The effect of this uncertainty can be modeled by rewiring a certain fraction of protein interactions.
More precisely, with a probability $q$, each reported interaction is ignored and a new interaction is randomly drawn
between two proteins that do not interact according to the available data. In this way we obtain a new network with a
certain degree of dissimilarity, depending upon $q$, with the original one. The degree of dissimilarity $f_l$ is measured
as the percentage of edges between proteins couples which are different in the two networks, the original and the
scrambled. Note that moving one link in general implies that the connectivity pattern of four vertices is affected and
that $f_l$ has a non trivial dependence on $q$. We implemented our method on the modified network, determining a new list
of putative functions for each unclassified protein, together with the relative probability (or frequency) of occurrence
of the putative functions themselves. For convenience we imagine these lists (one for each unclassified protein) to
contain all possible functions, and associate a zero probability to those functions that have never occurred in the
implementation of the method. We call $p_{is}(f_l)$ the probability that the unclassified protein $i$ belongs to the
functional class $s$, in the network with a degree of dissimilarity $f_l$ with the original one. The case $p_{is}(0)$
then corresponds to the functional classification obtained using the original network. A quantitative comparison with the
predictions made using the original network is provided by the overlap function $\Theta_i(f_l)$ defined as follows:  
$\Theta_i(f_l)=\sum_s\sqrt{p_{is}(0)p_{is}(f_l)}$, that equals 1 when $p_{is}(f_l)=p_{is}(0)$ for all $s$. We computed
the average of $\Theta_i(f_l)$ restricted to unclassified proteins with $k$ interacting partners, obtaining that it does
not vary too much with the node degree. In Fig. 2B we plot the average of $\Theta_i(f_l)$ over all unclassified proteins
as a function of $f_l$. For 10\% of dissimilarity, the overlap is around 0.85\%. Since each displaced edge corresponds to
three to four proteins with different interactions, this shows that even if about 30-40\% of proteins have at least a
misplaced interaction due to experimental erroneous results, an effective evaluation of the proteins' functions is not
precluded.  Of course larger levels of errors lower the overlap, signaling that the two networks provide rather different
configurations of functional assignment.

The method we propose appears as a general tool for the assignment of protein function pointing out that protein-protein
interaction data can be an effective framework to deduce the function of unclassified proteins. The method also allows to
determine multiple functionalities and takes into account self-consistently the effect of unclassified proteins in the
final assignment configuration. Finally, the validity tests performed show that the method tolerates the inherent
imperfection and the incomplete nature of the protein networks.

\section*{Experimental protocol}

To each unclassified protein $i=1,2, ... ,n$ a function si is assigned, chosen among the $F$ possible ones in order to
globally minimize the following score function: 

\begin{equation} 
E = - \sum_{i,j} J_{ij} \delta(\sigma_i,\sigma_j) - \sum_i h_i(\sigma_i)\ , \label{e} 
\end{equation} 

\noindent where $J_{ij}$ is the adjacency matrix of the interaction network for the unclassified proteins ($J_{ij}$ is
equal to 1 if protein $i$ and $j$ interact and are unclassified, 0 otherwise), $\delta(i,j)$ is the discrete delta
function and $h_i(\sigma_i)$ is the number of classified partners of protein $i$ with function $\sigma_i$.

The "majority rule" of [5,6] corresponds to minimize solely the second term on the r.h.s. of equation (1). The above
can be achieved with local methods (i.e. considering successively and independently each protein). In our method, the
contribution to the total score of assigning a protein $i$ to functional class $\sigma_i$ depends also on the assignment
made to all other proteins, resulting in a much harder computational task. The advantage is that the underlying
requirements that "interaction requires a common function" is applied also to interactions between previously
unclassified proteins, that are completely ignored with the majority rule approach.

To overcome the computational difficulties and find the configuration or configurations that minimize $E$ we perform a
simulated annealing [16] introducing an effective temperature $T$.  We start with an initial random configuration
$\sigma_i$.  Then, at each Monte-Carlo step, we select one protein at random and change its state from si to
$\sigma^\prime_i$, where $\sigma^\prime_i$ is selected at random among the possible states of protein $i$ with the
constraint $\sigma^\prime_i\neq\sigma_i$. We then compute the score difference $\Delta E=E^\prime-E$ between these
two configurations.  If $\Delta E\leq0$ we accept the new configuration. If $\Delta E>0$, we accept the new configuration
with probability $r=exp(-\Delta E/T)$ or keep the original configuration with probability $1-r$. This Monte-Carlo step is
repeated until $E$ reaches a stationary value. Thereafter, $T$ is decreased by a tiny amount. These two process,
equilibration at a given $T$ and decrease of $T$ is repeated until the protein states remain unchanged for a sufficiently
long time. At the end the protein states are our predicted functional classification.  Since the the minimum energy
solution is not unique the simulated annealing process has been repeated several times, starting from different initial
configurations.  Finally we computed the fraction of times pis the protein i was observed in the final state s, which
give us an estimate of the probability that protein i belongs to the functional classification s. References

\end{document}